\documentclass[a4paper,11pt]{article}
\usepackage{jinstpub} 
\usepackage{lineno}
\usepackage{siunitx}
\usepackage{cleveref}
\usepackage{textgreek}
\usepackage[utf8]{inputenc}
\usepackage[T1]{fontenc}

\title{\boldmath Performance Optimizations and Evaluations for the Small Direct Currents Measurement System}

\author[a]{Shunyi Liang}
\author[b,1]{Juncheng Liang\note{Co-corresponding authors.}}
\author[a,1]{Kezhu Song}
\author[a]{Yijie Jiang}
\author[b]{and Zhijie Yang}

\affiliation[a]{State Key Laboratory of Particle Detection and Electronics,\\ 
University of Science and Technology of China,\\
Hefei 230026, China}
\affiliation[b]{National Institute of Metrology,\\
Beijing 100029, China}
\emailAdd{liangjc@nim.ac.cn}
\emailAdd{skz@ustc.edu.cn}

\abstract{
Ionization chambers are essential for activity determinations in radionuclide metrology. 
We have developed a high-precision integrating-differentiating (int-diff) system for measuring small currents.
It is anticipated to enhance the ionization current measurement capability of the 4\textpi\textgamma~ionization chamber radioactivity standard at the National Institute of Metrology (NIM), China.
Besides, it has broad application prospects in physical experiments and fundamental metrology.
The design of the measurement system is optimized through circuit analysis and simulation.
The structure of the integrating capacitor array is redesigned to reduce the error of the amplification gain, 
and a relay is used as the reset switch to achieve improved noise and leakage performance.
The digital readout and control module is also enhanced in terms of flexibility and functionality.
High-precision test platforms utilizing the standard small current source at NIM China and an ionization chamber were developed to evaluate the performance of the system.
The results demonstrate an ultra-low noise floor (\SI{<1}{\femto\ampere/\surd\hertz}) and a low current bias of \SI{}{\femto\ampere}-level,
as well as a low temperature coefficient of the amplification gain of \SI{2.1}{ppm/\degreeCelsius}.
The short-term stability and linearity of the gain are also tested and exhibit comparable indicators to those of the Keithley 6430.
Reasonable results are obtained in the long-term reproducibility test.
Therefore, the system enables high-precision measurements for small direct currents and shows promise for applications in ionization chambers.
}


\begin{document}
\maketitle
\flushbottom

\section{Introduction}
\label{sec:intro}
Accurate measurement of small currents ranging from femtoamperes to nanoamperes is essential in radionuclide metrology.
In the process of activity determinations, a factor that often restricts the uncertainty lies in the measurement of the ionization current\cite{ptb_improved_2021}.
In recent years, picoampere-meters (picoammeters) have attained higher levels of precision and have been widely applied in various fields, including fundamental metrology~\cite{drung_improving_2015}, physical experiments~\cite{torre_rhip_2016}, and biosensors\cite{crescentini_noise_2014}.
In the National Institute of Metrology (NIM), China, researchers are dedicated to improving the 4\textpi\textgamma~ionization chamber (IC) radioactivity standard, thereby raising the demand for high-precision picoammeters.

The ultra-stable low-noise current amplifier (ULCA), based on the feedback trans-impedance amplifier (TIA) structure, was introduced by the Physikalisch-Technische Bundesanstalt (PTB). 
Its key element, the resistor array, provides very stable amplification gain, enabling achievement of the state-of-the-art uncertainty of 0.1 ppm in \SI{100}{\pico\ampere}~\cite{drung_ultrastable_2015}. 
The well-known commercial electrometers Keithley model 6430 and Keysight model B2980 both employed TIA to amplify small currents, and they have multiple ranges and functionalities to support different scenarios~\cite{sobolewski_source_2000, noauthor_b2980b_nodate}.
The TIA structure is relatively mature and precise,and it's widely applied in recent researches~\cite{wang_high_2013, mortuza_pico-current_2017, utrobicic_floating_2015}.
However, limited by the precision resistor techniques, large resistors providing high gain are basically less precise and less stable.
In comparison with another structure known as the shunt amplifier, the TIA structure can be more precise for its low input burden voltage, but the shunt type is more commonly seen in portable multi-meters because its simpler circuit and lower cost~\cite{epure_low-cost_2013, torre_rhip_2016}.
Certain studies adopted the integrating-differentiating (int-diff) structure to achieve the least noise level by replacing the TIA's noisy feedback resistor with a capacitor in the integrating stage (integrator)~\cite{crescentini_noise_2014, noauthor_axon_2021}.
Other than ultra-low noise performance, this structure can also accomplish high bandwidth that more than \SI{100}{\kilo\hertz}~\cite{crescentini_compact_2017}.
However, its drawbacks are as follows: the capacitor significantly influences the precision and stability of the gain as well as the analog differentiating stage can introduce nonlinearity errors.
Thus previous researches on this kind of amplifiers seldom took into account the stability and precision for DC measurements~\cite{crescentini_noise_2014}.

In our previous work~\cite{liang_low-noise_2023}, we proposed a precise small currents measurement system (SCMS). 
This system is based on the int-diff method to achieve low noise. 
It integrates the direct current (dc) in the analog amplification circuit and differentiates it in a field programmable gate array (FPGA) to restore a dc output.
The analog circuit combines careful circuit design and techniques. 
For instance, by means of the averaging effect of the integrating capacitor array, we took stability into considerations.

In this paper, we further optimized the design of the SCMS through analysis and simulation, and conducted more comprehensive tests.
The capacitor array structure was optimized through simulations to suppress gain errors due to parasitic capacitance.
The leakage current performance of the SCMS was analyzed and optimized.
We replaced the CMOS-based switches with a reed relay, and observed improved noise and leakage performance.
Additionally, the host software of the SCMS is upgraded to enhance its functionality.
We also developed different test platforms utilizing the standard small current source equipment at NIM China and a 4\textpi\textgamma~IC to evaluate various performance indicators of the SCMS.
The test results exhibit an ultra-low noise floor (\SI{<1}{\femto\ampere/\surd\hertz}), a \SI{}{\femto\ampere}-level current bias, a low temperature coefficient (\SI{2.1}{ppm/\degreeCelsius}), short-term stability and linearity comparable to those of the Keithley 6430, as well as reasonable long-term reproducibility for an IC application.

The paper is organized as follows: 
In Section \ref{sec:struct}, the architecture of the SCMS is introduced, with a primary focus on the optimizations made to its design.
Section \ref{sec:pt} presents the test setups devised for evaluating noise and stability performances and provides an analysis of the test results.
Section \ref{sec:con} concludes the paper.

\section{Measurement System Architecture}
\label{sec:struct}

\begin{figure}[t]
    \centering
    \includegraphics[width=3.5in]{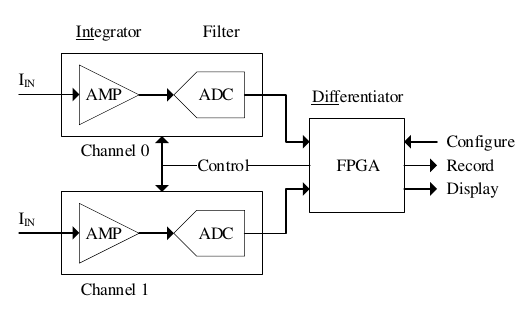}
    \caption{Overall architecture of the SCMS.\label{fig:meas-sys}}
\end{figure}

In our previous work \cite{liang_low-noise_2023}, we put forward a precise small currents measurement system. 
It has a basic structure as shown in \cref{fig:meas-sys}, including the analog front-end circuit and the digital readout and control module.
The system is extensible to support the simultaneous reading of two front-end circuits,
thereby improving the signal-to-noise ratio by measuring the current twice.
The measurement system is designed with an integrating-differentiating (int-diff) scheme,
One of our innovations is the analog-integrating and digital-differentiating method.
Since we don't use an analog differentiating circuit, 
such as a conventional correlated double sampling (CDS) circuit that can be an error source for the sample and hold circuit \cite{bennati_sub-pa_2009,baumgartner_systems_2006},
our structure ensures high precision and flexibility in the data processing.

\subsection{Front-End Circuit Design}

\begin{figure}[t]
    \centering
    \includegraphics[width=3.5in]{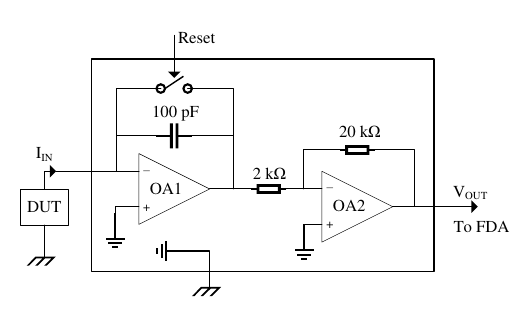}
    \caption{Simplified schematic of the analog amplification circuit.\label{fig:scms-schm}}
\end{figure}

The most critical part of this design is the front-end amplification circuit.
It comprises two stages, namely the low-noise integrating stage (integrator circuit),
and the voltage amplification stage. 
\Cref{fig:scms-schm} illustrates the basic structure of the two-stage amplification circuit.
The first stage integrates current to voltage ramp through a capacitor, 
and the second stage provides 10-fold inverting voltage gain.
The slope of the output signal is related to the measured input current with an equivalent trans-resistance gain $G_{\text{TR}}$ of
\begin{equation}
\label{eq:gain}
    G_{\text{TR}} = (- \frac{T_{\text{i}}}{C_{\text{i}}}) \times
        (- \frac{R_{\text{f}}}{R_{\text{i}}}) ,
\end{equation}
where $T_{\text{i}}$ and $C_{\text{i}}$ are the integrating time and \SI{100}{\pico\farad} integrating capacitor,  and $R_{\text{f}}$ of \SI{20}{\kilo\ohm} along with $R_{\text{i}}$ of \SI{2}{\kilo\ohm} are the feedback and input resistors of the second stage.
The integrator needs to be periodically reset by a switch to ensure that the output voltage will not saturate.

\begin{figure}[t]
    \centering
    \includegraphics[width=3.5in]{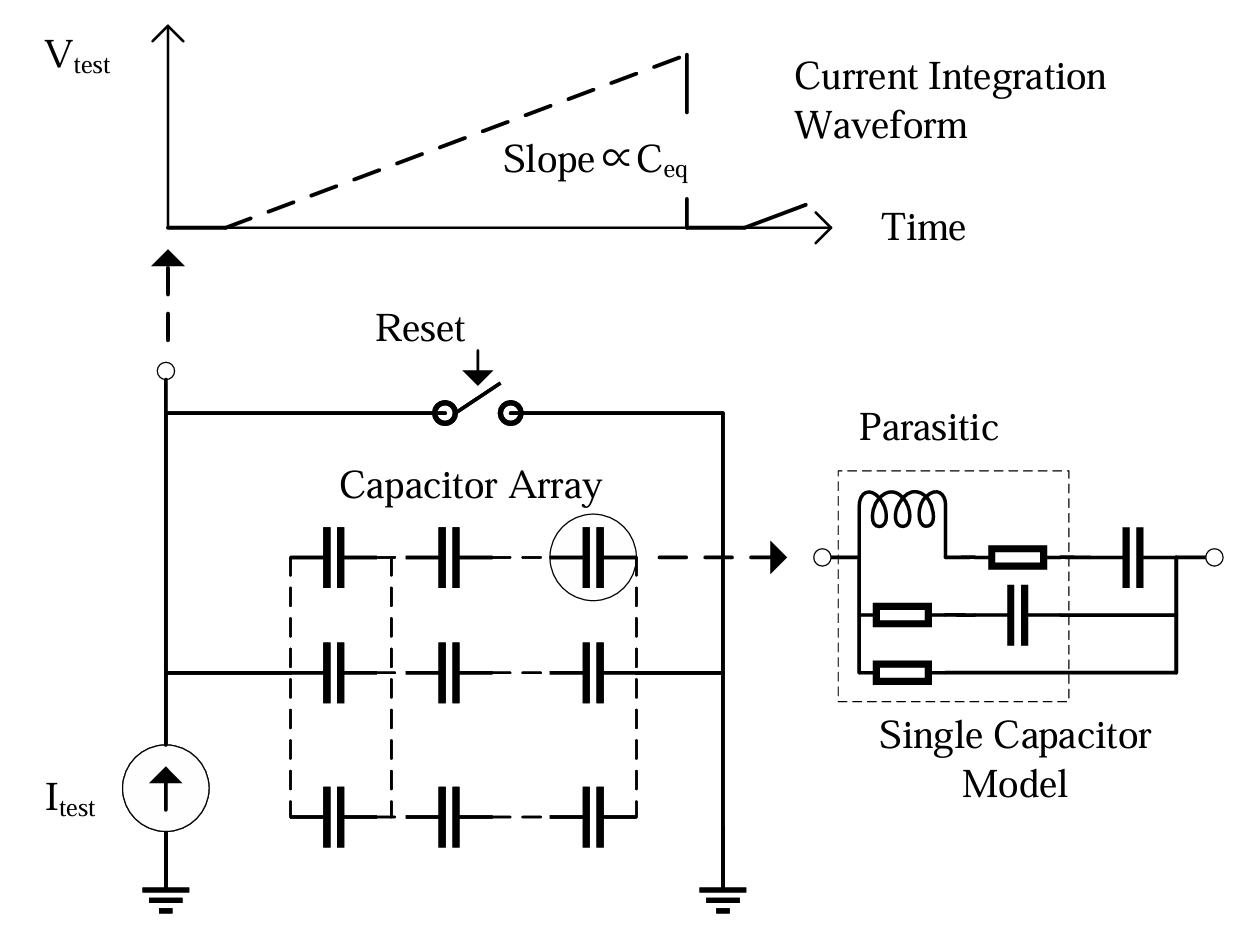}
    \caption{Illustration of the parasitic capacitance modeling circuit, assuming an ideal reset switch and an ideal testing current source, with multiple configurations for connecting the capacitor array. The equivalent capacitance is assessed by analyzing the slope of the current integration waveform.\label{fig:cas}}
    \end{figure}

To increase the insulation resistances (IR) and improve the precision and stability,
100 capacitors are arranged in serial and parallel configuration to form the equivalent \SI{100}{\pico\farad} integrating capacitor. 
The IR can be simply regarded as being connected in parallel with the integrating capacitor, thereby forming an $RC$ circuit
and resulting in an integrating error relative to the first stage gain\cite{liang_low-noise_2023},
\begin{equation}
    \delta_\text{int} = n(1-e^{-1/n})-1,
\end{equation}
where $n=R_\text{IR} C_\text{i}/ T_\text{i}$, and $R_\text{IR}$ represents the total IR.
Each capacitor is a class I ceramic capacitor (provided by KEMET) with more than \SI{100}{\giga\ohm} IR, low tolerance and low temperature coefficient.
A high IR not only decreases $\delta_\text{int}$ but also minimized the leakage current ($I_\text{L}$) 
according to $I_\text{L} = V_\text{C} / R_\text{IR}$,
where $V_\text{C}$ represents the voltage applied to the capacitor array.
We improved the structure of the integrating capacitor array in comparison to the previous 100 serial-connected capacitor array.
To account for the more intricate parasitics of the capacitors,
we constructed dedicated simplified circuit models simulating and analyzing the capacitor array performance, 
as illustrated in \cref{fig:cas}.

The simulations involved three main steps: netlist editing, circuit solving, and data analysis. Circuit solving was conducted using a simulation program with integrated circuit emphasis (SPICE), while Python was employed to enhance the other steps, especially for simulating complex circuits~\cite{salvaire_circuit_2017}.
The capacitor array was evaluated from the tolerance of the equivalent capacitance ($C_{\text{eq}}$) and the nonlinearity error of the current integrating waveform.
Firstly, tolerance is calculated as the relative deviation between the measured and nominal values of $C_{\text{eq}}$. 
We conducted Monte Carlo simulations, where the capacitor's value follows a normal distribution.
Secondly, the nonlinearity error $\delta_\text{int}$ is determined as the maximum relative residual of the linear fit to the integration waveform. 

Simulations revealed that tolerance and nonlinearity error are influenced by the array structure and capacitor selection, necessitating comprehensive considerations.
Table~\ref{tab:ca-sim} presents the performance of various capacitor arrays, each identified by an abbreviated label indicating its structure. For instance, 'p$X$-s$Y$' denotes $X$ capacitors in parallel, forming a block, with $Y$ blocks connected in series.
The results demonstrate that the 's5-p5-s4' structure, utilizing \SI{430}{\pico\farad} 0603 capacitors, achieves optimal performance.

\begin{table}[t]
    \centering
    \caption{Simulated Performance Comparison of Various Capacitor Array Configurations.\label{tab:ca-sim}}
    \smallskip
    \begin{tabular}{|l|l|l|c|c|}
    \hline
    structure & capacitor & $C_{\text{eq}}$ & tolerance & nonlinearity \\
    \hline
    s10-p10-s4 & \SI{430}{\pico\farad} 0805 & \SI{107.5}{\pico\farad} & \SI{0.068}{\%} & \SI{39.4}{ppm}\\
    s5-p5-s4 & \SI{430}{\pico\farad} 0603 & \SI{107.5}{\pico\farad} & \SI{0.022}{\%} & \SI{6.8}{ppm}\\
    s100 & \SI{10}{\nano\farad} 0603 & \SI{100}{\pico\farad} & \SI{-0.53}{\%} & \SI{31.6}{ppm}\\
    s20-p20 & \SI{100}{\pico\farad} 0603 & \SI{100}{\pico\farad} & \SI{0.702}{\%} & \SI{180}{ppm}\\
    s10-p10-s4 & \SI{430}{\pico\farad} 0603 & \SI{107.5}{\pico\farad} & \SI{0.039}{\%} & \SI{7.42}{ppm}\\
    \hline
    \end{tabular}
\end{table}
    
A parasitic capacitance at the pF level paralleled to the integrating capacitor is observed, 
which directly affects the equivalent capacitor precision.
The PCB parasitic capacitance was evaluated and optimized through post-layout simulation using Cadence PowerSI. 
It is calculated in \eqref{eq:ec} based on the simulated impedance ($Z_{\text{C}}$) at low frequencies. 
\begin{equation}
    \label{eq:ec}
        C_{\text{p}} = \frac{1}{2\pi f Z_{\text{C}}}.
    \end{equation}
We optimized the PCB layout by arranging the capacitor array tightly, 
removing the connector and raising the height between the signal and reference layers.
Simulation result indicates that the parallel parasitic capacitance decreases from \SI{8}{\pico\farad} to \SI{3}{\pico\farad}.

Leakage current is a key factor influencing the precision of ultra-low current measurements;
thus, efforts are required to minimize it.
It can be generated from several components of the first-stage amplifier circuit, 
such as the operational amplifier (OA), the capacitor array, the reset switch, the PCB layout, and the input cabling.
The ADA4530-1 was selected as the first-stage OA, which features a bias current as low as \SI{3}{\femto\ampere} (typical).
The guarding technique was adopted in the PCB design to reduce leakage by enclosing the measurement path with a low-impedance shield\cite{noauthor_keithley_nodate}. 
After soldering, 
the PCB was carefully cleaned with isopropyl alcohol (IPA) in an ultrasonic cleaner 
to remove any potentially conductive contaminants\cite{vicky_wong_ada4530-1_2015}.
Low-noise triaxial cables with double-layer shielding were applied in the tests to minimize the triboelectric effect and cable leakage.
Furthermore, using the method of controlling variables, we removed each component in the circuit one by one and measured the leakage current (bias current).
Eventually, we found that the CMOS-based switch contributed a leakage current at the \SI{100}{\femto\ampere} level, 
which was much higher than that of other circuit components.

To minimize leakage,
the reset switch was enhanced by substituting the CMOS-based switch with a signal reed relay.
Although the CMOS-based switch offers high switching speed, 
the signal relay (provided by Panasonic) exhibits ultra-low leakage current and low inherent noise characteristics, as verified in \cref{sec:pt}.
The drawbacks of the relay switch include a lower switching speed and significant signal fluctuations within approximately \SI{100}{\milli\second} after the switch is opened (current integrating phase).
The fluctuation may be caused by the fact that this relay is not shielded, 
and the switch control signal is coupled into the current trace.
We solved this issue by employing long integrating-reset period of \SI{0.5}{\second},
and discarding the first \SI{100}{\milli\second} data after the switch is opened.

The second-stage OA uses the ADA4625 with balanced noise and offset performance.
The resistors are metal foil resistors with ultra-low temperature coefficient and high accuracy.
A fully differential amplifier (FDA) circuit is inserted between the amplification circuit and the analog-to-digital converter (ADC). It realizes single-ended to differential conversion and implements a multi-feedback second-order Bessel filter with \SI{10}{\kilo\hertz} bandwidth.
The FDA circuit drives a high-precision successive approximation register (SAR) type ADC, 
which has a high signal-to-noise ratio, high linearity and low distortion, facilitating accurate signal sampling.

\subsection{Read-out and Control Design}

\begin{figure}[t]
    \centering
    \includegraphics[width=3.5in]{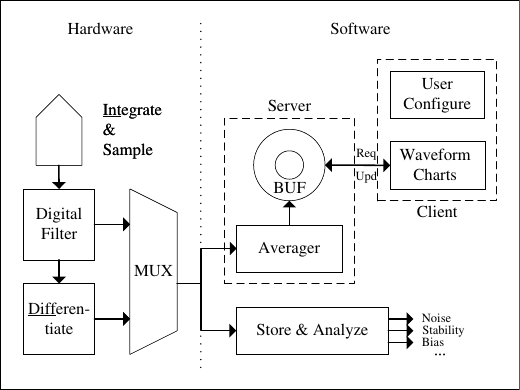}
    \caption{Architecture of the SCMS. It involves hardware amplification and digitization of the input current, with subsequent data analysis performed in software.\label{fig:ms}}
\end{figure}

The readout and control module consists of an FPGA-based digital circuit and host computer software. 
Its structure is illustrated in \cref{fig:ms}.
The digital circuit is designed to be isolated from the front-end through \SI{100}{\mega\hertz} high-speed digital isolators.
This design disrupts the ground loop of the analog and digital circuits, 
thus reducing the interference and noise affecting the sensitive analog circuit.
The analog circuit is independently powered by two batteries and a battery management system (BMS), in order to achieve low-noise power supply, and it is covered by two-layer shieldings.

The FPGA logic is designed to be flexible, and supports two-channel multi-ports data readout.
Port A is the output data of a SINC1 digital filter, which exhibits less overshoot and higher waveform accuracy.
This data can be utilized for system debugging as it has high data readout rate (\SI{23}{\kilo SPS}) and represents the amplified voltage waveform.
The port A data is then differentiated by a differentiating logic to convert integrated waveform into a dc output (port D).
This is implemented by subtracting two adjacent data points, and the differentiation process is in real-time.
The digital circuit is designed to be flexibly configurable,
enabling software access to both the filtered waveform and the differentiated measurements.

The data is transmitted using a universal serial bus (USB) interface to the host software.
The system's software is developed in Python, enabling online user configuration, data storage, displaying, and offline analysis.
For online display and configuration, the system employs a web browser/server (B/S) architecture to facilitate remote multi-user access.
Within the browser client, users can configure the display settings and hardware parameters for sampling and filtering.
The server uses a repeat average filter to combine data points per $n$ power line cycles (PLC), 
 thus reducing noise glitches and decreasing the data rate.
The processed data is stored in a ring buffer, periodically extracted by the user client to update waveform charts.
Additionally, the software supports comprehensive offline analysis with detailed analysis methods outlined in Section \ref{sec:pt}.

\section{Performance Test Platforms and Results}
\label{sec:pt}

Our prior research evaluated the system's performance by testing its noise level,
output offsets, and transient response\cite{liang_low-noise_2023}.
However, some crucial specifications, such as the stability of the amplification gain and the inherent current bias and noise,
which significantly impact precision, could not be tested using the existing setup.
Therefore, the platforms and methods for precisely testing these performance are investigated.

\subsection{Current Bias and Noise Test}

In this test, the current input termination of the SCMS was left open-floating and shielded with a metal cap.
We continuously sampled data for several hours, filtered it, and calculated the noise spectral density (NSD) and the Allan deviation (ADEV) offline.

To research the influence of the environment temperature, a built-in temperature sensor is employed to record the temperature of the system.
The temperature sensor is placed near the integrating capacitor array.
We tested in both thermostatic and non-thermostatic environments.
The fluctuations of the current bias and temperature for both tests are listed in \cref{tab:bias_curr}.
Since the current bias represents the current measured when the input is disconnected, it reflects the leakage performance.
The results demonstrate that the optimized SCMS exhibits a low leakage at the \SI{}{\femto\ampere}-level.

\begin{table}[t]
    \centering
    \caption{Current Bias Measurement Results\label{tab:bias_curr}}
    \smallskip
    \begin{tabular}{lcc}
        \hline
        temperature range                    & current bias range                   & averaged current bias     \\
        \hline
        \SIrange{26.8}{27.1}{\degreeCelsius} (thermostatic) & \SIrange{-0.19}{0.62}{\femto\ampere} & \SI{0.24}{\femto\ampere}  \\
        \SIrange{26.9}{27.9}{\degreeCelsius} (non-thermostatic) & \SIrange{-2.91}{0.01}{\femto\ampere} & \SI{-1.02}{\femto\ampere} \\
        \hline
    \end{tabular}
\end{table}

\begin{figure}[t]
    \centering
    \includegraphics[width=4.7in]{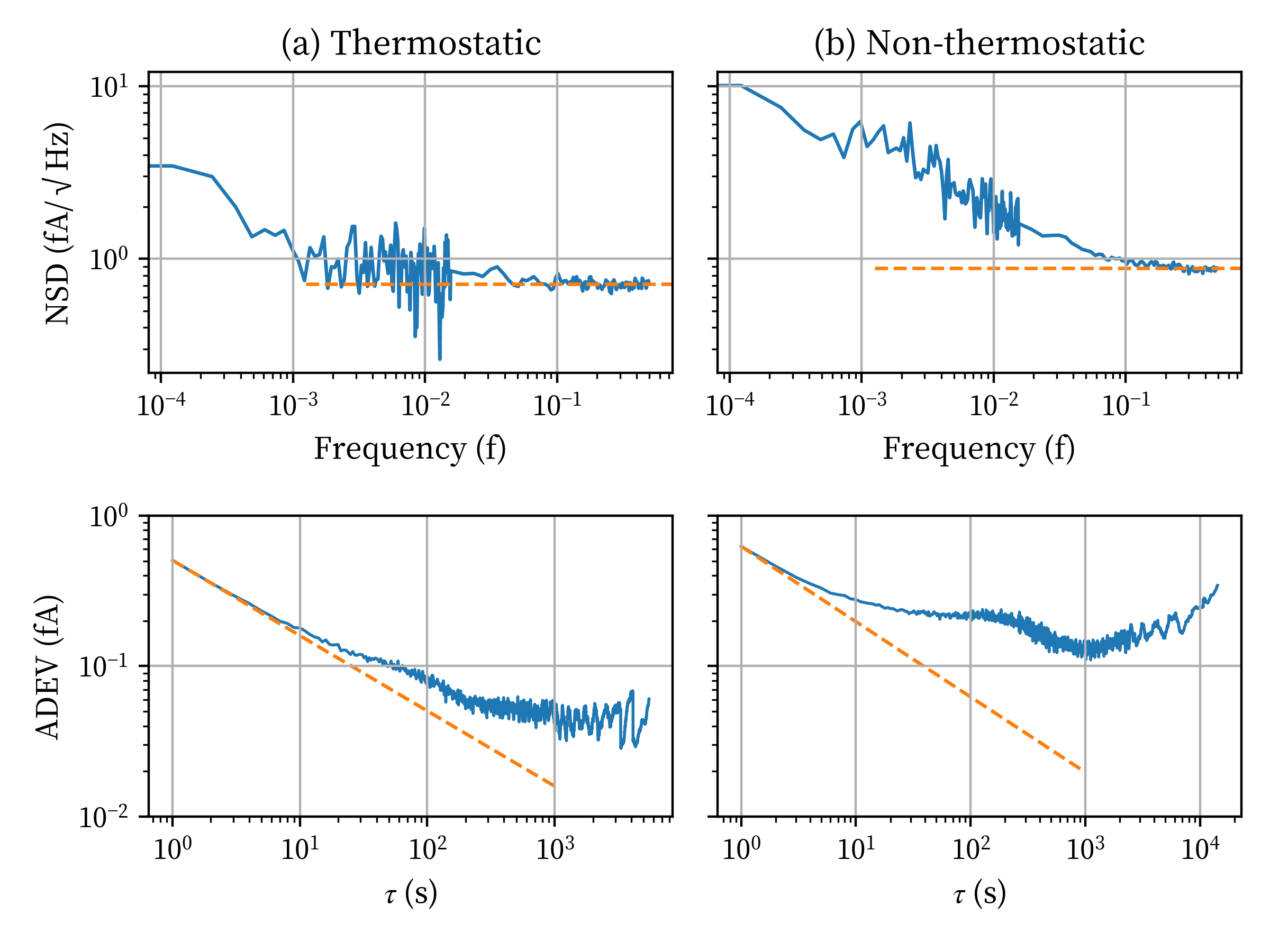}
    \caption{Noise test results analyzed by NSD and ADEV.\label{fig:noise}}
\end{figure}

\Cref{fig:noise} depicts the noise performance of the system,
and the current measurement data is the same as the bias current test.
Results indicate that in both thermostatic and non-thermostatic environments,
the system has a low white noise level of \SI{0.7}{\femto\ampere\surd\hertz} and \SI{0.9}{\femto\ampere\surd\hertz}.
In DC measurements, consecutive data sampling and averaging are commonly used for improving precision.
ADEV is widely used for characterizing the noise and fluctuations in low-level measurement instruments \cite{scherer_electrometer_2019,witt_using_2001}.
The slow fluctuations of the temperature induced low-frequency current fluctuations,
and elevated the ADEV in non-thermostatic condition.
The lowest ADEV, which is also called "bias stability",
is about \SI{0.04}{\femto\ampere} in thermostatic environment.
ADEV can be used to evaluate the impact of both white noise and low-frequency noise on data variability under different integrating time $\tau$.
The orange straight lines represent the white noise contributions.
In ADEV plots, they are calculated by
\begin{equation}
    \sigma_\text{w} = \sqrt{S_\text{w}/2\tau},
\end{equation}
where $S_\text{w}$ is the white NSD\cite{rutman_characterization_1991}.
Compared with our previous design that employed CMOS-based reset switch (with an current bias of \SI{100}{\femto\ampere} level and a noise level of \SI{7}{\femto\ampere\surd\hertz})\cite{liang_low-noise_2023},
the noise and bias performance have been significantly improved.

\subsection{Stability Tests}

\begin{figure}[t]
    \centering
    \includegraphics[width=3.5in]{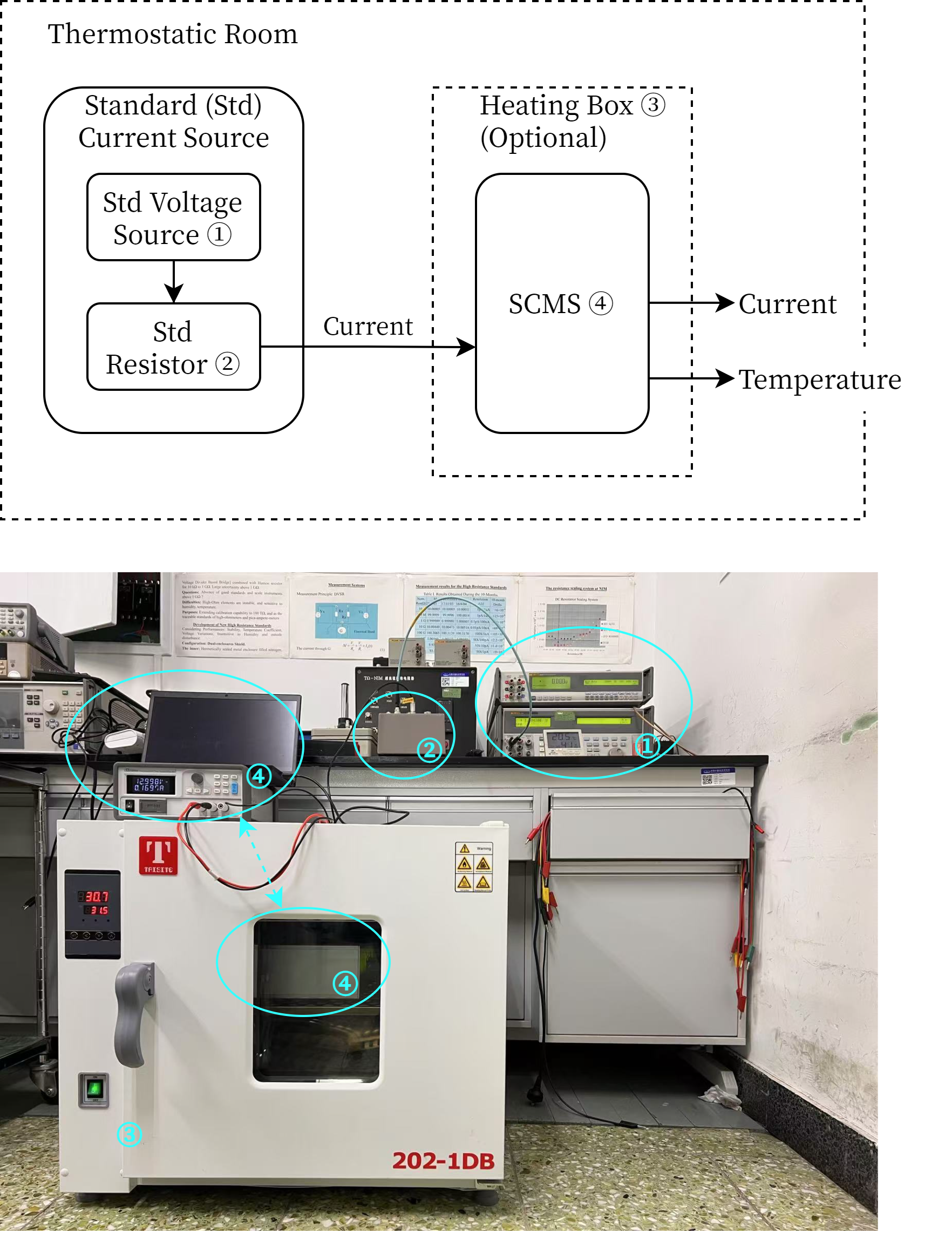}
    \caption{Stability tests setup. The heating box is optional, and dedicated to be used in temperature dependency test.\label{fig:stab-test}}
\end{figure}

The stability performances of the SCMS gain includes short-term stability, temperature dependency, linearity and long-term reproducibility.
A test platform is built utilizing the standard small current source equipment at NIM China, as shown in \cref{fig:stab-test}.
The standard current source comprises a standard voltage source (Fluke 5520A) and a standard resistor (Keithley 5156).
The standard equipment is calibrated regularly at NIM China, and the key parameters are presented in \cref{tab:volt-src,tab:res}.
The current NSD of Keithley 5156 is estimated using resistor thermal noise equation
\begin{equation}
    I_\text{N} = \sqrt{4 k T / R},
\end{equation}
where k is the Boltzmann constant, T is the Kelvin temperature and R is the resistance.
This standard equipment is the most precise calibrator currently available.

\begin{table}[t]
    \centering
    \caption{Fluke 5520A Standard Voltage Source Parameters}
    \smallskip
    \label{tab:volt-src}
    \begin{tabular}{lcccl}
        \hline
        parameters                & \SI{330}{\milli\volt} range     & \SI{3.3}{\volt} range              & \SI{33}{\volt}range               & note                                \\
        \hline
        peak-peak noise           & \SI{1}{\micro\volt}             & \SI{10}{\micro\volt}               & \SI{100}{\micro\volt}             & \SIrange{0.1}{10}{\hertz}           \\
        equivalent voltage NSD    & \SI{53}{\nano\volt/\surd\hertz} & \SI{0.53}{\micro\volt/\surd\hertz} & \SI{5.3}{\micro\volt/\surd\hertz} & estimated as white noise            \\
        short-term stability      & 3ppm+\SI{1}{\micro\volt}        & 2ppm+\SI{1.5}{\micro\volt}         & 2ppm+\SI{15}{\micro\volt}         & 24 hours, \SI{\pm1}{\degreeCelsius} \\
        long-term reproducibility & 15ppm+\SI{1}{\micro\volt}       & 9ppm+\SI{2}{\micro\volt}           & 10ppm+\SI{20}{\micro\volt}        & 90 days, \SI{\pm5}{\degreeCelsius}  \\
        \hline
    \end{tabular}
\end{table}

\begin{table}[t]
    \centering
    \caption{Keithley 5156 Standard Resistor Parameters}
    \label{tab:res}
    \smallskip
    \begin{tabular}{lcccc}
        \hline
        parameters               & \SI{1}{\giga\ohm}                    & \SI{10}{\giga\ohm}                   & \SI{100}{\giga\ohm}                  \\
        \hline
        current NSD              & \SI{4.07}{\femto\ampere/\surd\hertz} & \SI{1.29}{\femto\ampere/\surd\hertz} & \SI{0.41}{\femto\ampere/\surd\hertz} \\
        temperature coefficient & <\SI{25}{ppm/\degreeCelsius}         & <\SI{25}{ppm/\degreeCelsius}         & <\SI{100}{ppm/\degreeCelsius}        \\
        voltage coefficient     & <\SI{1}{ppm/\volt}                   & <\SI{1}{ppm/\volt}                   & <\SI{1}{ppm/\volt}                   \\
        \hline
    \end{tabular}
\end{table}

The testing uncertainties of the short-term tests (including temperature dependency and linearity tests) rely on the short-term stability of the standard equipment.
Since our primary focus is on the relative fluctuations of the gain,
the standard equipment calibration error, as a systematic error, will not affect the test results.
It comprises the stability of the voltage source (shown in \cref{tab:volt-src}) and the resistor.
The resistor's stability mainly stems from the temperature coefficient.
As we observed, in a single test, the temperature fluctuation is less than \SI{0.5}{\degreeCelsius}.
Therefore, the short-term stability of the resistor is considered to be 13 ppm of \SI{1}{\giga\ohm} and \SI{10}{\giga\ohm}, and 50 ppm of \SI{100}{\giga\ohm}.
The total relative error of this standard equipment is calculated under the worst-case scenario (where the correlation coefficient between the voltage and resistor is -1), and the formula is
\begin{equation}
    \label{eq:ua}
    \delta_\text{max} = |\delta_\text{u}| + |\delta_\text{r}|,
\end{equation}
where $\delta_\text{u}$ and $\delta_\text{r}$ are the relative stability errors of the voltage source and the resistor.
This is regarded as the limiting error, so the coverage factor $k$ is taken as 3.

To suppress low-frequency noise and drift effects\cite{scherer_electrometer_2019},
the currents are periodically reversed (every \SI{600}{\second}), and the first \SI{50}{\second} samples after current reversal are discarded to allow the current to stabilize.
We define the normalized gain ($G_\text{nom}$) as the measured current value $I_\text{m}$ divided by the standard source current $I_\text{s}$, as shown below:
\begin{equation}
    G_\text{nom} \coloneqq \frac{I_\text{m}}{I_\text{s}}.
\end{equation}
$I_\text{m}$ and current bias $I_\text{b}$ can be computed from the positive and negative current measurements ($I_\text{m+}$ and $I_\text{m-}$) as follows:
\begin{align}
    I_\text{m} & = \frac{1}{2}\left( I_\text{m+} - I_\text{m-} \right),                \\
    I_\text{b} & = \frac{1}{2}\left( I_\text{m+} + I_\text{m-} \right) / G_\text{nom}.
\end{align}
Each measurement corresponds to the averaged result obtained over a current reversion period of \SI{1200}{\second}.

\begin{figure}[t]
    \centering
    \includegraphics[width=3.5in]{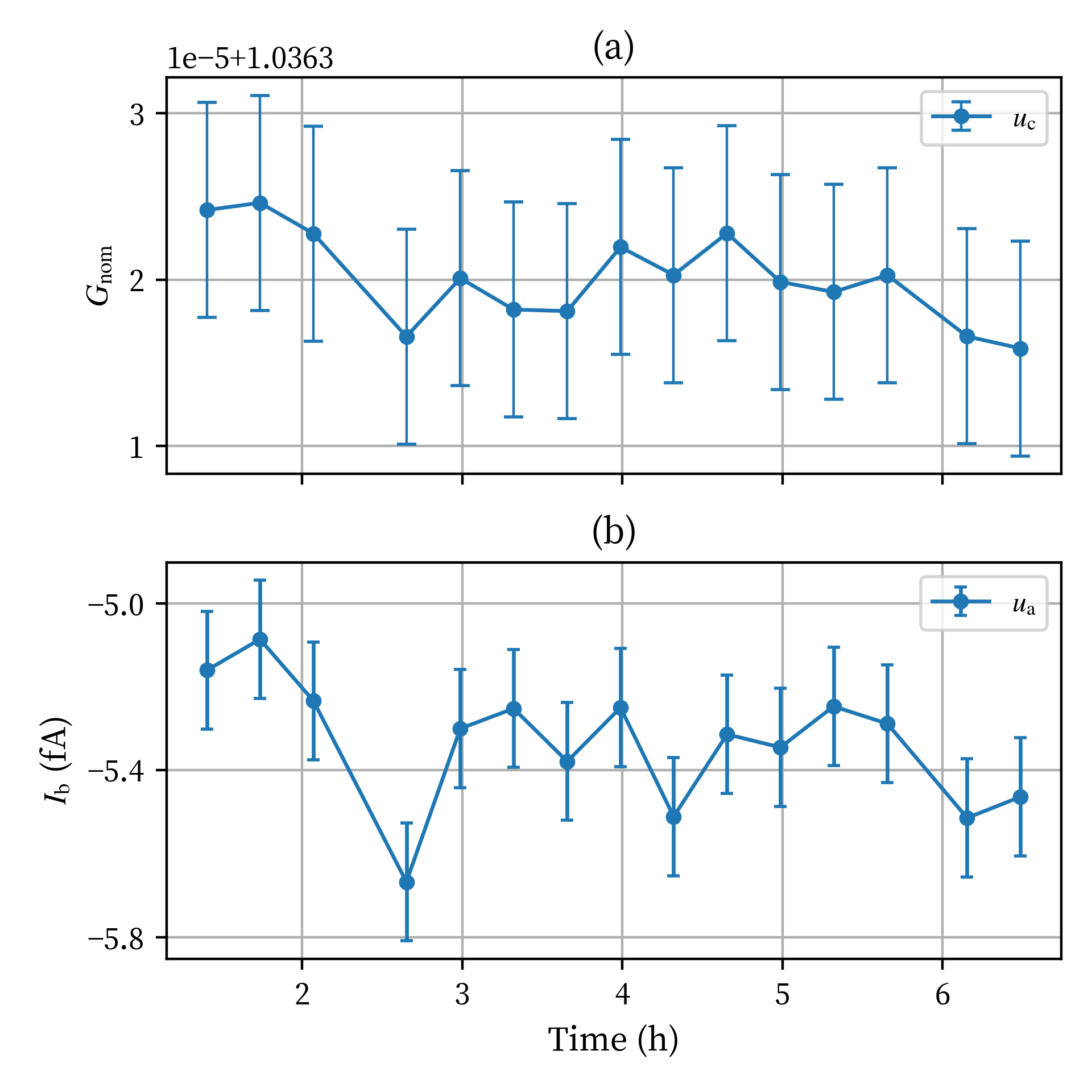}
    \caption{Short term stability test result. Here, $u_\text{c}$ represents the standard combined uncertainty,
    while $u_\text{a}$ stands for the Type A uncertainties.\label{fig:short-term}}
\end{figure}

During the short-term stability test, the standard equipment used the \SI{10}{\giga\ohm} resistor and the \SI{1}{\volt} voltage output to generate a \SI{100}{\pico\ampere} current.
\Cref{fig:short-term} shows the fluctuations of both the normalized gain $G_\text{nom}$ and current bias $I_\text{b}$ during the short-term stability test.
The bias current is the sum of that of the SCMS and the standard equipment.
Based on \cref{tab:bias_curr}, it is believed that the standard equipment plays a dominant role in the bias current result.

The ADEV is utilized to evaluate the data variability of the $G_\text{nom}$, the calculation formula is as follows:
\begin{equation}
    \label{eq:avar_est}
    \sigma_y = \sqrt{ \frac{\Sigma_{k=1}^{N-1}
            \left[ I_\text{m}(k+1) - I_\text{m}(k) \right]^2 }
        {2 \cdot (N-1)}},
\end{equation}
where $N$ is the number of the measurements.
The ADEV can be considered as a Type A standard uncertainty ($u_\text{a}$), which amounts to \SI{1.9}{ppm}.
The Type B standard uncertainty ($u_\text{b}$) is calculated based on the short-term stability of the standard equipment in \cref{eq:ua} and is \SI{6.2}{ppm}.
In total, the standard combined uncertainty ($u_\text{c}$) of the normalized gain is \SI{6.5}{ppm}.
All uncertainties are presented in the form of relative values.
We found that there is a 3.63\% deviation between the normalized gain and its nominal value of 1.
The deviation is attributed to the parallel parasitic capacitance as previously analyzed.

The temperature coefficient of the standard current source equipment at NIM China is inadequate for
accurately evaluating the temperature dependency of the SCMS.
Therefore, we came up with an idea of placing the standard device in the thermostatic room to guarantee the stability of the test current,
while placing the SCMS inside a heating box to vary its temperature.
Before this test, the instruments had been preheated for one hour.
Moreover, after each temperature alteration, we waited for one hour to allow the temperature to stabilize.

\begin{figure}[t]
    \centering
    \includegraphics[width=3.5in]{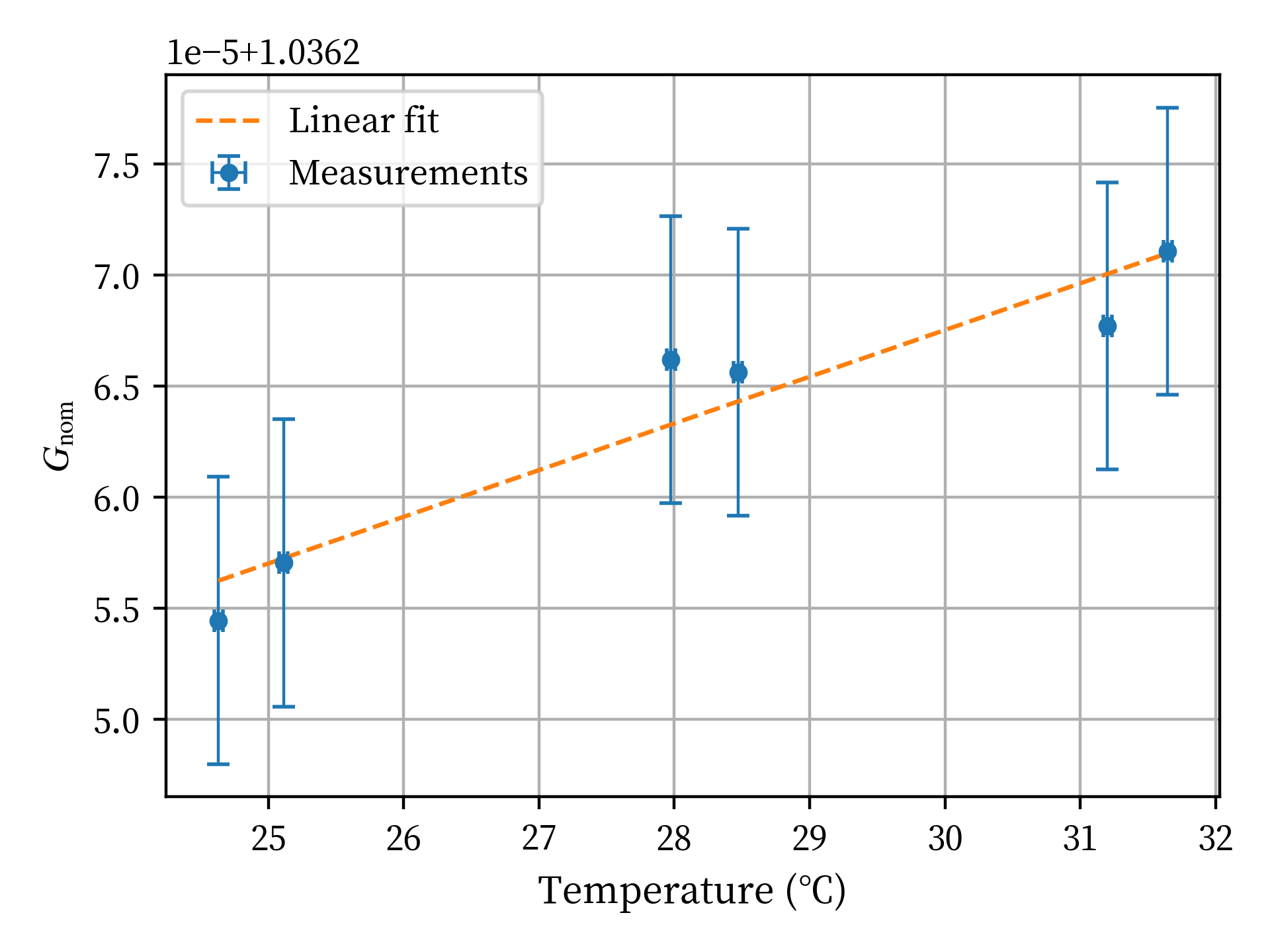}
    \caption{Temperature dependency test result.\label{fig:temp}}
\end{figure}

The test result is shown in \cref{fig:temp}.
There is an obvious correlation between the gain and temperature,
and the coefficient of determination ($R^2$) for the linear fit is 0.91.
According to the linear regression analysis, the slope represents the temperature coefficient of the SCMS,
which is about $2.1\pm 1 $ \SI{}{ppm/\degreeCelsius}.
The integrated temperature sensor of the SCMS has a very low reproducibility error of \SI{32}{\milli\degreeCelsius}, so that it's negligible.

For the linearity test, the standard resistor is set to \SI{100}{\giga\ohm}, and we adjust the standard voltage output to obtain a small current ranging from \SI{10}{\femto\ampere} to \SI{100}{\pico\ampere}.
We use the fixed resistor to mitigate the impact of calibration errors of different resistors.
$u_\text{a}$ in this test is the same as that in the short-term stability test,
and $u_\text{b}$ of different current values are estimated using the same method.
In fact, $u_\text{a}$ can be lower since we use a larger resistor that generates less thermal noise.
The standard uncertainties corresponding to each current measurement are listed in \cref{tab:lnrt}.

\begin{table}[t]
    \centering
    \caption{Current Values and Uncertainties in the Linearity Test}
    \label{tab:lnrt}
    \smallskip
    \begin{tabular}{ccccc}
        \hline
        input current & voltage & $u_\text{c}$ \\
        \hline
        \SI{100}{\pico\ampere}  & \SI{10}{\volt}        &  \SI{19}{ppm}   \\
        \SI{30}{\pico\ampere}   & \SI{3.0}{\volt}       &  \SI{20}{ppm}   \\
        \SI{10}{\pico\ampere}   & \SI{1.0}{\volt}       &  \SI{27}{ppm}   \\
        \SI{3.0}{\pico\ampere}  & \SI{300}{\milli\volt} &  \SI{68}{ppm}   \\
        \SI{1.0}{\pico\ampere}  & \SI{100}{\milli\volt} &  \SI{200}{ppm}  \\
        \SI{300}{\femto\ampere} & \SI{30}{\milli\volt}  &  \SI{650}{ppm}  \\
        \SI{100}{\femto\ampere} & \SI{10}{\milli\volt}  &  \SI{0.19}{\%}  \\
        \SI{30}{\femto\ampere}  & \SI{3.0}{\milli\volt} &  \SI{0.65}{\%}  \\
        \SI{10}{\femto\ampere}  & \SI{1.0}{\milli\volt} &  \SI{1.9}{\%}   \\
        \hline
    \end{tabular}
\end{table}

\begin{figure}[t]
    \centering
    \includegraphics[width=3.5in]{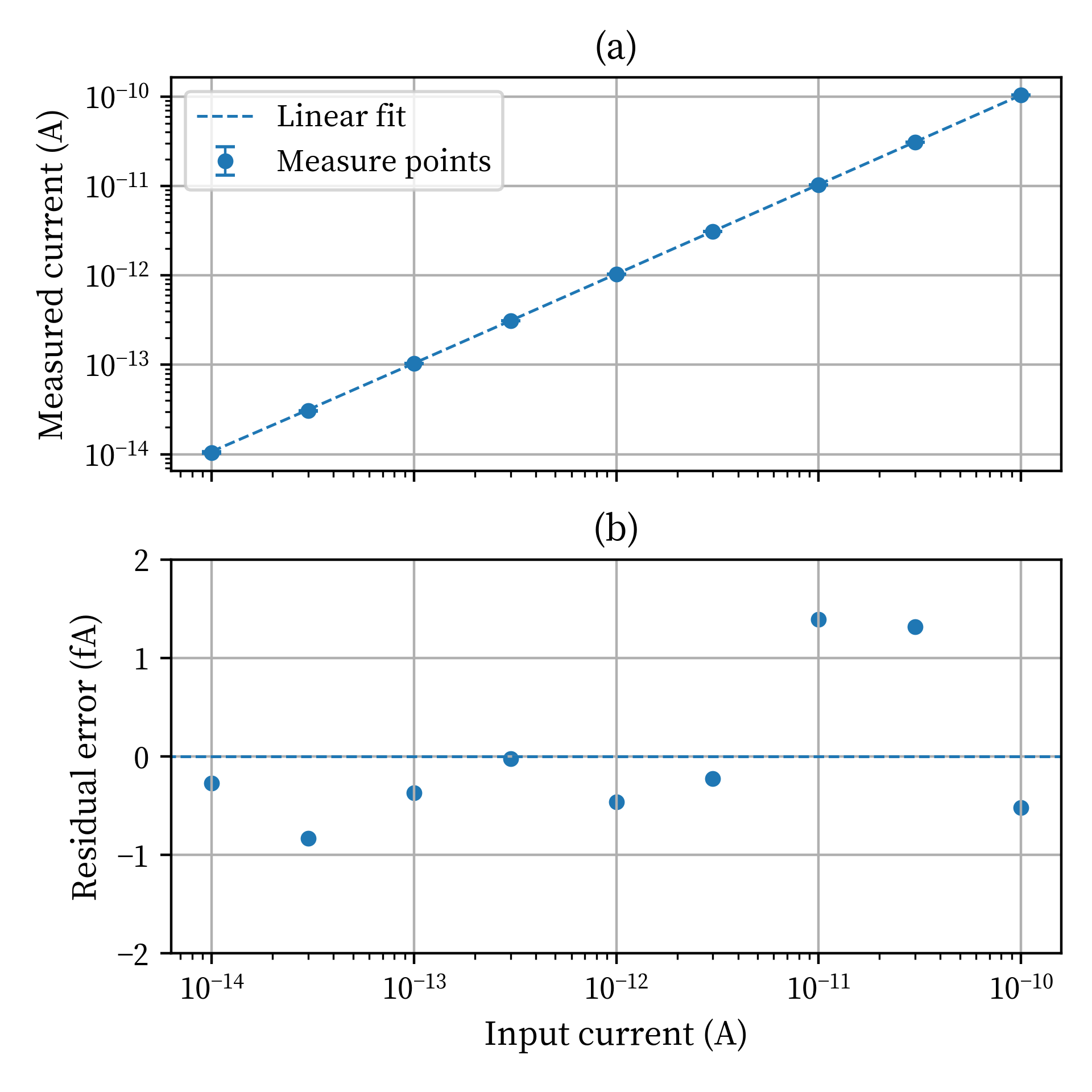}
    \caption{Linearity test result.\label{fig:lnrt}}
\end{figure}

The linear test result is shown in \cref{fig:lnrt}.
The measured current and the input current are highly correlated, with a coefficient of determination ($R^2$) greater than 0.999,999,999.
The residual errors are plotted in subfigure (b). The maximum residual error is \SI{1.4}{\femto\ampere} and the standard deviation of the residuals is \SI{0.86}{\femto\ampere}.

\begin{figure}[t]
    \centering
    \includegraphics[width=3.5in]{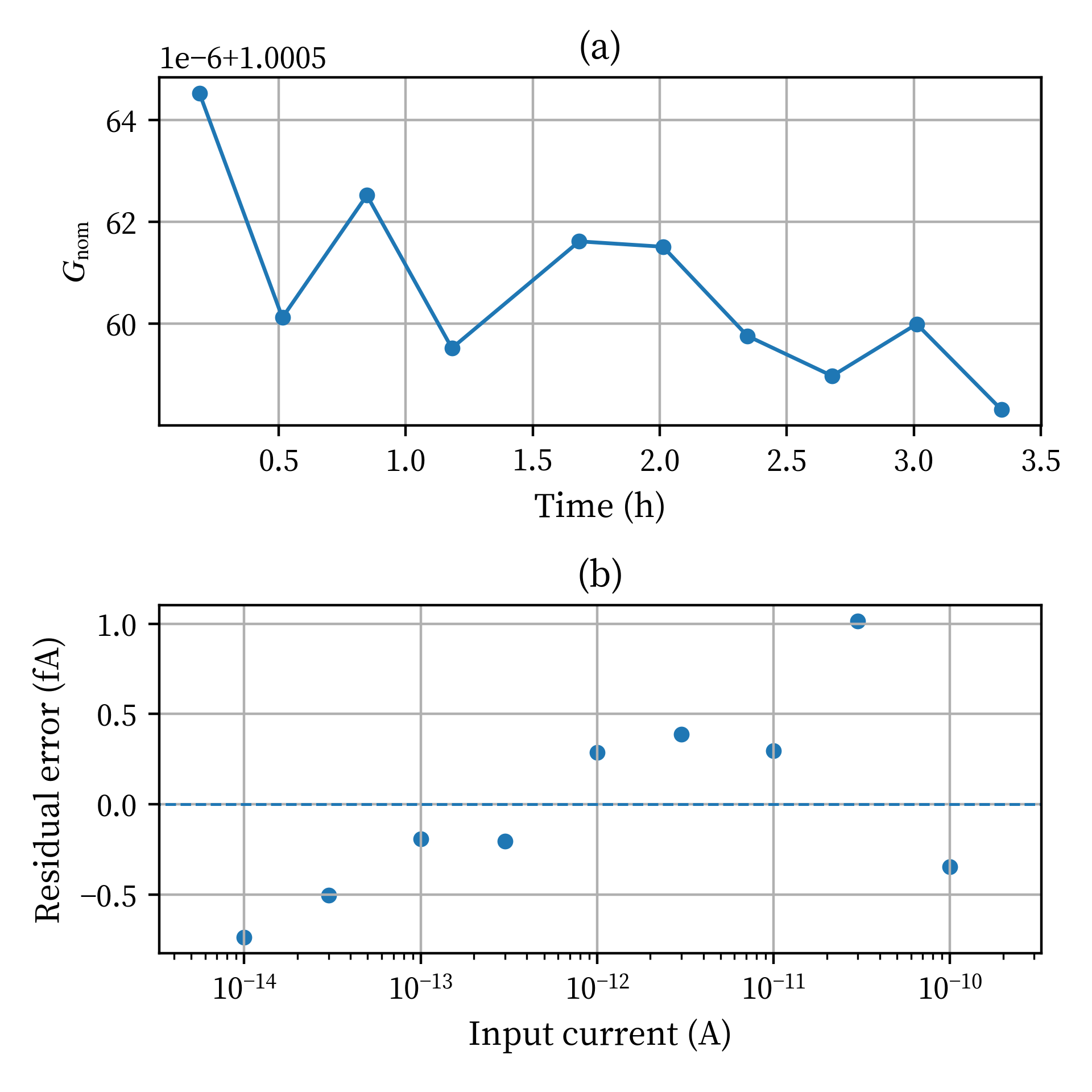}
    \caption{Stability test results of Keithley 6430 for comparison.\label{fig:k6-lnrt}}
\end{figure}

Utilizing the same test platform, we assessed the linearity and short-term stability of a commercial digital picoammeter, the Keithley 6430.
The \SI{100}{\pico\ampere} range along with the highest-accuracy mode of this instrument was selected for comparison with the SCMS,
and the moving average and median average filters were disabled.
The results of short-term stability and linearity tests are presented in \cref{fig:k6-lnrt},
The ADEV with a \SI{1200}{\second} integration time in the short-term test is \SI{1.6}{ppm},
and the standard deviation of the residuals is \SI{0.58}{\femto\ampere}.
In these two tests, Keithley 6430 and the SCMS exhibit similar performances.
However, as its temperature coefficient, which we calculated from the datasheet, is at the \SI{100}{ppm/\degreeCelsius} level, it is much larger than that of the SCMS.

The input voltage burden $V_\text{b}$ of the SCMS was also tested.
We used a nanovoltmeter (Keithley 2182) to measure the voltage difference between the current input (through a buffer) and the ground.
For both the input current of \SI{\pm 100}{\pico\ampere} and \SI{\pm 10}{\pico\ampere}, the $V_\text{b}$ remained stable at around \SI{33}{\micro\volt}. 
For comparison, the $V_\text{B}$ of Keithley 6430 is larger and exhibits instability in response to variations in the input current \cite{scherer_electrometer_2019}. 

\subsection{Long-term Reproducibility Test Utilizing Ionization Chamber}

\begin{figure}[t]
    \centering
    \includegraphics[width=3.5in]{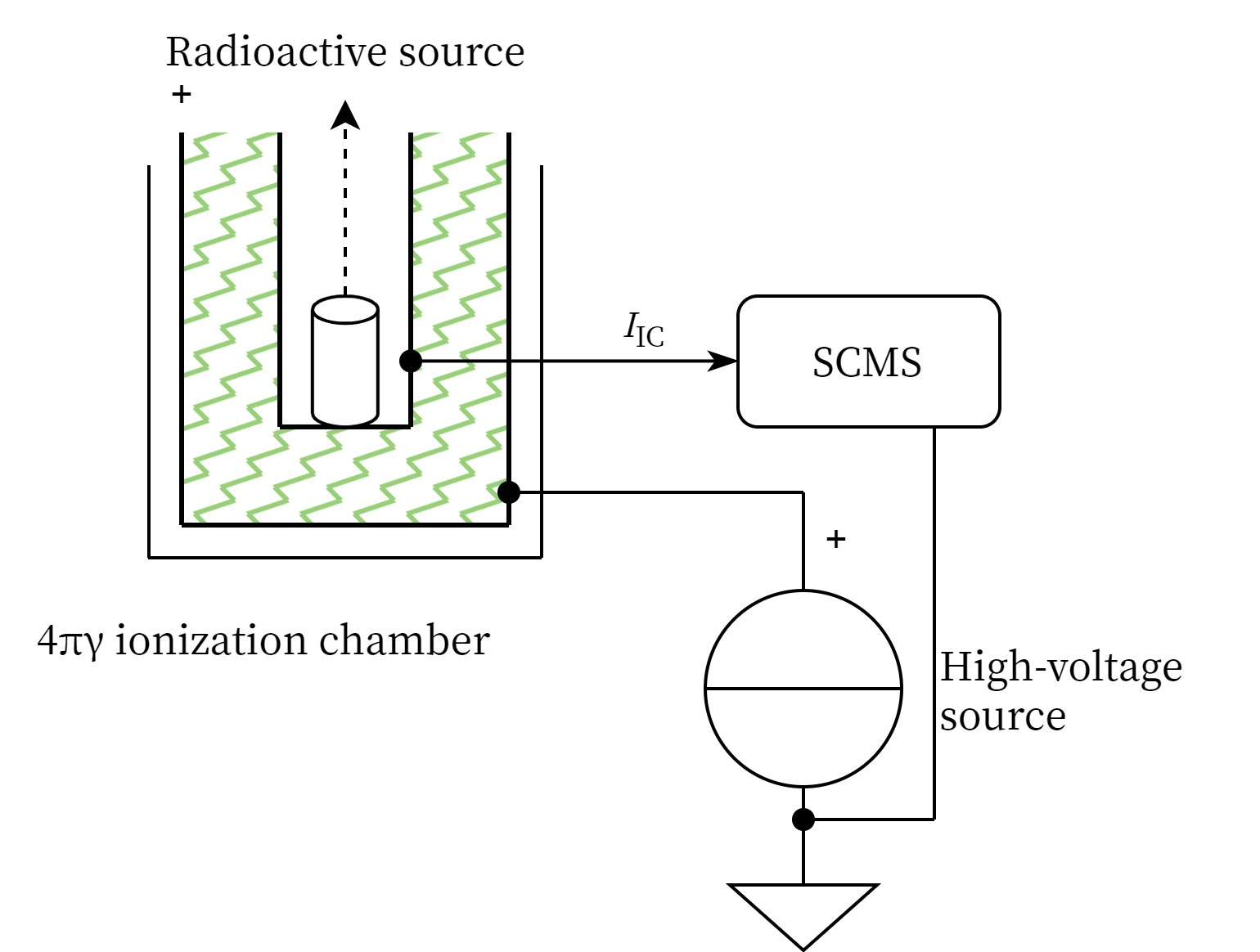}
    \caption{Ionization chamber test platform.\label{fig:ic-setup}}
\end{figure}

Since the standard current source equipment at NIM China cannot be occupied for a long time, 
the long-term stability of the SCMS was tested using an IC and a radioactive source with a long half-life.
The IC is a 4\textpi\textgamma~ionization chamber manufactured in China,
and it's connected to the SCMS through a \SI{0.3}{\meter}-long low-noise triaxial current cable.
A voltage of \SI{600}{\volt} is applied to provide the polarization electric field to the IC.
The test setup is illustrated in \cref{fig:ic-setup}.

The noise performance when using an IC is more complex than that of the SCMS itself.
This is because the equivalent capacitance model of the IC increases the noise gain of the amplification circuit, and the high-voltage source introduces excess noise\cite{giblin_exploring_2019,cottini_minimum_1956}.
In this research, several ICs and high-voltage sources were combined to conduct a preliminary assessment of the NSD, following which the combination with the lowest noise was selected.
A \textsuperscript{226}Ra radioactive source was employed, which has a long half-life of approximately 1600 years.
Thus, during the one-month testing period, the attenuation is only \SI{36}{ppm} and can be considered negligible.
During the test, the test platform was placed in an air-conditioned room.
On average, the tests were carried out twice a week.
For each measurement, a preheating process lasting \SI{1.5}{\hour} was performed first, followed by a \SI{30}{\min} measurement of the IC's background current ($I_\text{bg}$), and then a \SI{30}{\min} measurement of the ionization current $I_\text{ic}$.
The background-free measurement result is calculate as follows:
\begin{equation}
    I_\text{m} = \bar{I}_\text{ic} - \bar{I}_\text{bg}.
\end{equation}
The duration of a single measurement is kept within \SI{1}{\hour} to reduce the error caused by the background current fluctuation\cite{fitzgerald_next_2020}.

\begin{figure}[t]
    \centering
    \includegraphics[width=3.5in]{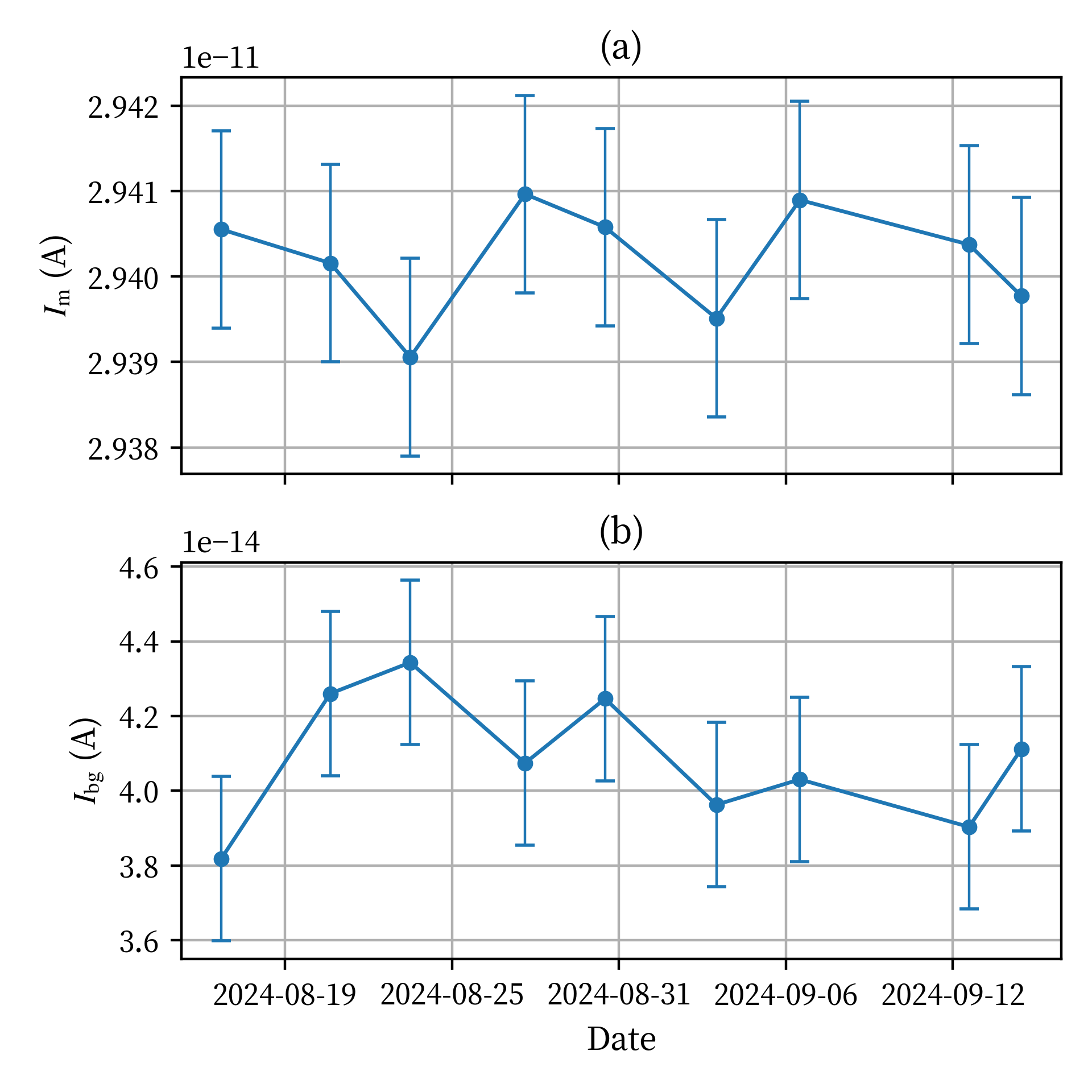}
    \caption{Long-term reproducibility test results. The error bars represent the $u_\text{a}$.\label{fig:ltp}}
\end{figure}

The results of the long-term reproducibility test are shown in \cref{fig:ltp}.
To evaluate the $u_\text{a}$ of this test, 
we continuously repeated the measurements ($I_\text{m}$) several times after preheating.
And then we used the difference between two adjacent measurements to calculate the ADEV by applying \cref{eq:avar_est}.
The ADEV for $I_\text{m}$ and $I_\text{bg}$ tests are \SI{12}{\femto\ampere} and \SI{2.2}{\femto\ampere}, respectively, 
and the averaged values of $I_\text{m}$ and $I_\text{bg}$ during this test are \SI{29.4}{\pico\ampere} and \SI{40.8}{\femto\ampere}.
Due to the complexity of the $u_\text{b}$ in the IC test, we do not discuss it in this study.

\section{Conclusions}
\label{sec:con}
A high-precision measurement system for pA level small direct currents has been developed.
It is intended to replace the commercial picoammeter and be applied to the 4\textpi\textgamma~IC radioactivity standard at NIM, China,
thereby improving the ionization current measurement capability.
Meanwhile, it can serve as a high-precision digital picoammeter, which has broad application prospects in the fields of physical experiments and fundamental metrology.
The SCMS has been optimized on its gain stability and noise performance.
Through simulations, the configuration of the capacitor array was optimized to mitigate the impact of parasitic capacitance on gain.
Meanwhile, the primary source of noise and current leakage has been identified and minimized by using a relay as the reset switch.
Furthermore, the digital readout and control module has been upgraded to enhance its flexibility and functionalities.

The test setups for noise and stability were conducted using the standard small current source equipment and the IC.
The SCMS exhibits a low noise floor of less than \SI{1}{\femto\ampere\surd\hertz} and a \SI{}{\femto\ampere}-level current bias.
The low current bias demonstrates that the SCMS has been optimized to achieve ultra-low leakage performance.
Moreover, its noise performance is further improved compared with our previous research.
In terms of stability, the gain of the SCMS has a very low temperature coefficient (\SI{2.1}{ppm/\degreeCelsius}),
as well as short-term stability and linearity indicators similar to those of Keithley 6430.
The long-term reproducibility was also studied by using an 4\textpi\textgamma~IC, and reasonable test results were obtained.
In future work, we plan to conduct in-depth research on the electronic model and complex error sources of the IC, 
so as to further improve the measurement capability of the ionization current.

\acknowledgments

This work was supported by the NIM, China.
We would like to express our gratitude to Bo Liang from the Electromagnetic Metrology Division of NIM China,
for providing the test conditions and her valuable suggestions for this work.

\bibliographystyle{JHEP}
\bibliography{scms-jinst.bib}







\end{document}